\documentclass[aps,pra,showpacs,superscriptaddress,amssymb,twocolumn]{revtex4-1}
\usepackage{graphicx}
\usepackage{appendix}
\usepackage{color}
\usepackage[caption=false]{subfig}
\usepackage{epstopdf} 
\DeclareGraphicsExtensions{.pdf,.eps,.png,.jpg,.mps}
\usepackage[colorlinks=true,citecolor=blue,urlcolor=blue,linkcolor=blue]{hyperref}
\usepackage{bm,amsmath}
\usepackage{dcolumn}

\newcommand{\ket}[1]{\left|#1\right\rangle}

\def\BEq{\begin{equation}}
\def\EEq{\end{equation}}
\def\BEqA{\begin{eqnarray}}
\def\EEqA{\end{eqnarray}}
\def\BW{\begin{widetext}}
\def\EW{\end{widetext}}

\begin{document}

%\title{Understanding leakage in superconducting error detection circuits}
\title{Understanding the effects of leakage in superconducting quantum error detection circuits}

\author{Joydip Ghosh}
\email{ghoshj@ucalgary.ca}
\affiliation{Institute for Quantum Science and Technology, University of Calgary, Calgary, Alberta T2N 1N4, Canada}
\affiliation{Department of Physics and Astronomy, University of Georgia, Athens, Georgia 30602, USA}
\author{Austin G. Fowler}
\email{austingfowler@gmail.com}
\affiliation{Department of Physics, University of California, Santa Barbara, California 93106, USA}
\affiliation{Centre for Quantum Computation and Communication Technology, School of Physics, The University of Melbourne, Victoria 3010, Australia}
\author{John M. Martinis}
\email{martinis@physics.ucsb.edu}
\affiliation{Department of Physics, University of California, Santa Barbara, California 93106, USA}
\author{Michael R. Geller}
\email{mgeller@uga.edu}
\affiliation{Department of Physics and Astronomy, University of Georgia, Athens, Georgia 30602, USA}

\date{\today}

\begin{abstract}

The majority of quantum error detection and correction protocols assume that the population in a qubit does not leak outside of its computational subspace. For many existing approaches, however, the physical qubits do possess more than two energy levels and consequently are prone to such leakage events. Analyzing the effects of leakage is therefore essential to devise optimal protocols for quantum gates, measurement, and error correction. In this work, we present a detailed study of leakage in a two-qubit superconducting stabilizer measurement circuit. We simulate the repeated ancilla-assisted measurement of a single $\sigma^z$ operator for a data qubit, record the outcome at the end of each measurement cycle, and explore the signature of leakage events in the obtained readout statistics. An analytic model is also developed that closely approximates the results of our numerical simulations. We find that leakage leads to destructive features in the quantum error detection scheme, making additional hardware and software protocols necessary.

\end{abstract}


\maketitle

\section{Introduction}

In many approaches to quantum computing, the physical qubits contain more than two levels, and the computational $\ket{0}$ and $\ket{1}$ states are defined within a subspace of the full ``qubit" Hilbert space. For such realizations of physical qubits, there exists a finite probability that the population tunnels out of the computational subspace, a phenomenon often referred to as \emph{leakage} \cite{PhysRevLett.83.5385,PhysRevB.81.104505,1367-2630-13-6-065030}. Understanding the effects of leakage is important for superconducting qubits not only because  higher energy states $\ket{2}, \ket{3}, \dots$ are present \cite{You2005,Neeley07082009}, as is the case with most other qubit realizations, but also because they can be utilized to implement two-qubit entangling operations such as the controlled-$\sigma^{\rm z}$ ({\sf CZ}) gate \cite{PhysRevLett.91.167005,DiCarlo2009,PhysRevA.87.022309}. While the critical effects of leakage on single \cite{Motzoi2009} and two-qubit quantum gates \cite{PhysRevA.87.022309,2013arXiv1306.6894E} have already been explored, its role in fault-tolerant quantum computation still remains an active area of research \cite{0034-4885-76-7-076001,PhysRevA.86.062318,PhysRevX.2.041003,2013arXiv1308.6642F}.

In this work, we investigate the role of leakage in the context of quantum error detection, a problem that we regard as a first step towards a comprehensive understanding of its impact on fault-tolerant quantum computing. We consider a realistic model for two coupled superconducting qubits, each having three energy levels, and simulate the repeated ancilla-assisted measurement of a single $\sigma^z$ operator on a data qubit. The ancilla qubit is measured at the end of each measurement cycle and readout statistics are obtained after many such cycles. We regard any population transfer to the $\ket{2}$ state of the ancilla or data qubit as a potential error, investigate the origin and signature of such leakage errors, and discuss the consequences for quantum error correction. It is observed that a typical data-qubit leakage event manifests itself by producing a ``noisy" ancilla qubit that randomly reads $\ket{0}$ or $\ket{1}$ from cycle to cycle. Although the measurement operation is compromised, the presence of the leakage event is thus apparent and detectable. However, there is also the possibility of a less typical but more dangerous type of leakage event, where the ancilla becomes \emph{paralyzed}, rendering it oblivious to data-qubit errors for many consecutive measurement cycles and thereby compromising the entire error detection scheme. We identify certain dynamical phases associated with the entangling gate that determine which type of leakage event (regular or paralyzing) will occur in practice. Leakage errors occur in most qubit realizations and our model and results may be relevant for quantum error detection and correction protocols in architectures beyond those based on superconducting devices.

\begin{figure*}[htb]
\includegraphics[angle=0,width=\textwidth]{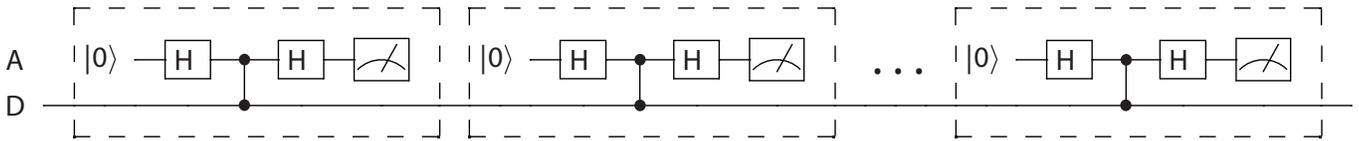}
\caption{Protocol for ancilla-assisted $\sigma^z$ measurement. Here ``A" is the ancilla qubit and ``D" the data qubit. Each cycle (dashed box) consists of a reset of the ancilla to the $\ket{0}$ state, a Hadamard gate {\sf H} on the ancilla, a {\sf CZ} gate, and another Hadamard followed by ancilla readout in the diagonal basis. The readout result is recorded and the cycle is repeated indefinitely. The data qubit never gets measured or reset.}
\label{fig:circuit1}
\end{figure*}

Figure \ref{fig:circuit1} shows the circuit for our protocol. Let's review how this works in the ideal limit: Initially, the data qubit D is assumed to be in some pure qubit (not qutrit) state 
\begin{equation}
\ket{\psi_{\rm D}}= a \ket{0}+ b \ket{1},
\label{initial data state}
\end{equation}
while the ancilla A is initialized to $\ket{0}$. We perform the gate operations shown in Fig.~\ref{fig:circuit1}, record the measurement outcome, reset the ancilla to $\ket{0}$, and repeat this cycle many times. Throughout this work, the Hadamard gate {\sf H} is assumed to be ideal and to act as the identity on the third level of the qutrit,
\begin{equation}
{\sf H} \equiv \left[\begin{array}{ccc}
 \frac{1}{\sqrt{2}} &  \frac{1}{\sqrt{2}} & 0\\
 \frac{1}{\sqrt{2}} &  -\frac{1}{\sqrt{2}} & 0 \\
 0 & 0 & 1 \\
\end{array}\right].
\label{H definition}
\end{equation}
The Hadamards and {\sf CZ} combine to produce a controlled-NOT ({\sf CNOT}) gate that copies the data qubit to the ancillla, but here we implement this {\sf CNOT} with the gates shown in Fig.~\ref{fig:circuit1} because theory predicts that the {\sf CZ} gate can be implemented in superconducting architectures with very high fidelity \cite{PhysRevA.87.022309}. For an initial state (\ref{initial data state}), the state of the system after the second {\sf H} gate, in the $| {\rm A} {\rm D}\rangle$ basis, is
\begin{equation}
a \ket{00}+ b\ket{11}.
\label{ideal final state}
\end{equation}
Thus, in the absence of any errors or decoherence, the readout projects the data qubit into the observed eigenstate of the ancilla. And once the data qubit is projected to a computational basis state, it remains there forever. 

In this work we study the effects of intrinsic gate errors and decoherence on this process. In Sec.~\ref{sec:CoupledQutritModel} we describe our physical model and consider ancilla-assisted measurement in the presence of decoherence. The non-ideal {\sf CZ} gate is discussed in Sec.~\ref{sec:parameterizationNICZ}. Leakage errors and ancilla paralysis are discussed in Sec.~\ref{sec:LeakageSignature}. We discuss the implications of our results for the design of error-corrected superconducting quantum computers in Sec.~\ref{sec:Conclusion}.

\section{Coupled qutrit model}
\label{sec:CoupledQutritModel}
 
In this section we describe our model, and for a warm-up, show how the ancilla-assisted measurement protocol works with ideal gates, but in the presence of decoherence.
 
\subsection{Model}

\begin{figure}[htb]
\includegraphics[angle=0,width=\linewidth]{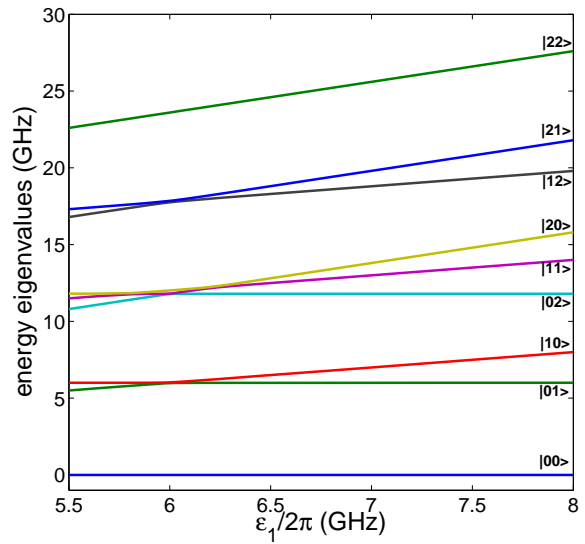}
\caption{(Color online) Energies of various levels, in the $\ket{\rm AD }$ basis, as a function of $\epsilon_1/2\pi$. Here $\epsilon_2/2\pi=6 \, {\rm GHz}$, the coupling strength is $g/2\pi=25 \, {\rm MHz}$, and $\eta_1/2\pi=\eta_2/2\pi=200$ MHz.}
\label{fig:energyLevels}
\end{figure}

The Hamiltonian for a pair of capacitively coupled transmon or phase qutrits is given by
\BEq
\label{eq:qutritHamiltonian}
H(t) = \left[\begin{array}{ccc}
 0 & 0 & 0 \\
 0 & \epsilon_1 & 0 \\
 0 & 0 & 2\epsilon_1-\eta_1 \\
 \end{array}\right]_{\rm \! \! q_{1}} + \left[\begin{array}{ccc}
 0 & 0 & 0  \\
 0 & \epsilon_2 & 0 \\
 0 & 0 & 2\epsilon_2-\eta_2 \\
 \end{array}\right]_{\rm \! \! q_{2}}+g Y \otimes Y,
\EEq
where
\BEq
Y \equiv \left[\begin{array}{ccc}
 0 & -i & 0\\
 i & 0 & -i\sqrt{2}\\
 0 & i\sqrt{2} & 0\\
 \end{array}\right].
 \label{Y definition}
\EEq
Qutrit 1 is the ancilla qutrit and qutrit 2 is the data qutrit. In (\ref{Y definition}) we have assumed harmonic qutrit eigenfunctions. The time-dependence of the Hamiltonian (\ref{eq:qutritHamiltonian}) is embedded in the qubit frequencies $\epsilon_1$ and $\epsilon_2$; the Hadamard gates are implemented with microwaves via terms not shown in (\ref{eq:qutritHamiltonian}).  For the {\sf CZ} gate protocol, we assume the frequency of the data qubit to be fixed at $6 \, {\rm GHz}$, while the ancilla's frequency is varied. The anharmonicities $\eta_i/2\pi$ are assumed to be equal, frequency-independent, and fixed at $200 \, {\rm MHz}$. Figure \ref{fig:energyLevels} shows the energies of several relevant eigenstates as a function of $\epsilon_1$, with $\epsilon_2/2\pi=6$ GHz and coupling strength $g/2\pi=25$ MHz. Note that the only avoided crossing at $\epsilon_1=\epsilon_2+\eta_1$ ($\epsilon_1/2\pi=6.2$ GHz in Fig.~\ref{fig:energyLevels}) is between the $\ket{11}$ and $\ket{20}$ channels; we use this anticrossing for our {\sf CZ} gate \cite{PhysRevLett.91.167005,PhysRevA.87.022309}. 

The {\sf CZ} gate, both ideal and non-ideal, is parameterized in this work via its {\it generator}. A generator of any unitary matrix $U$ is defined as a Hermitian matrix $S$ such that $U=e^{iS}$. For a two-qutrit system, the generator of the ideal {\sf CZ} gate is a Hermitian matrix $S$, whose matrix representation in the tensor-product basis
\begin{equation}
|{\rm AD}\rangle = 
 \big\lbrace \ket{00},\ket{01}\ket{02},\ket{10},\ket{11},\ket{12},\ket{20},\ket{21},\ket{22}  \big\rbrace
 \nonumber 
\label{two qutrit basis}
\end{equation} 
is
\BEq
\label{eq:SCZ}
S =\left[\begin{array}{ccccccccc}
 0 & 0 & 0 & 0 & 0 & 0 & 0 & 0 & 0 \\
 0 & 0 & 0 & 0 & 0 & 0 & 0 & 0 & 0 \\
 0 & 0 & \xi_1 & 0 & 0 & 0 & 0 & 0 & 0 \\
 0 & 0 & 0 & 0 & 0 & 0 & 0 & 0 & 0 \\
 0 & 0 & 0 & 0 & \pi & 0 & 0 & 0 & 0 \\
 0 & 0 & 0 & 0 & 0 & \xi_2 & 0 & 0 & 0 \\
 0 & 0 & 0 & 0 & 0 & 0 & \pi & 0 & 0 \\
 0 & 0 & 0 & 0 & 0 & 0 & 0 & \xi_3 & 0 \\
 0 & 0 & 0 & 0 & 0 & 0 & 0 & 0 & \xi_4 \\
 \end{array}\right].
\EEq
Note that within the computational subspace, $e^{iS}$ acts as a standard {\sf CZ} gate, while four of the five non-computational basis states acquire phases $e^{i\xi_i}$. We emphasize that any extension of an ideal {\sf CZ} gate to qutrits is dependent on the assumed model and gate protocol. For the Strauch {\sf CZ} gate, auxiliary $\sigma^z$ rotations on the ancilla and data qubits nullify the phases acquired by the $\ket{01}$ and $\ket{10}$ channels \cite{PhysRevA.87.022309,Mariantoni07102011}. Since we use the anticrossing between $\ket{11}$ and $\ket{20}$, they each acquire the same phase of angle $\pi$. We assume that the gate is in the adiabatic regime, and that the parameters $\xi_i$ are dynamical phases, which can then be expressed as
\begin{equation}
\begin{array}{l}
\displaystyle \xi_1 \approx -\int\limits_{0}^{t_{\rm gate}} \! \! E_{02} \, dt =
-\int\limits_{0}^{t_{\rm gate}}(2 \epsilon_2 - \eta_2) \, dt, \\
\displaystyle \xi_2 \approx -\int\limits_{0}^{t_{\rm gate}}  \! \! E_{12}  \,  dt =
-\int\limits_{0}^{t_{\rm gate}}(2 \epsilon_2 - \eta_2) \,
dt-\int\limits_{0}^{t_{\rm gate}} \! \epsilon_{1} \, dt, \\
\displaystyle \xi_3 \approx -\int\limits_{0}^{t_{\rm gate}}  \! \!  E_{21}  \, dt =
-\int\limits_{0}^{t_{\rm gate}} \! \epsilon_2  \, dt-\int\limits_{0}^{t_{\rm
gate}} \! \!  \left(2\epsilon_{1} -\eta_1\right) \, dt, \\
\displaystyle \xi_4 \approx -\int\limits_{0}^{t_{\rm gate}}  \! \!  E_{22}  \,  dt =
-\int\limits_{0}^{t_{\rm gate}}(2 \epsilon_2 - \eta_2) \,
dt-\int\limits_{0}^{t_{\rm gate}} \! \!  \left(2\epsilon_{1} -\eta_1\right) dt.
\\
\end{array} 
\label{eq:xdef}
\end{equation}
Here $t_{\rm gate}$ is the operation time for the {\sf CZ} gate (including auxiliary $z$ rotations), and $E_{ij}$ is the energy of eigenstate $\ket{ij}$, shown in Fig.~\ref{fig:energyLevels}. To keep our analysis general we do not assume specific values for the $\xi_i$. They depend on the details of the {\sf CZ} gate implementation but remain fixed throughout a given experiment or simulation (unless one changes $t_{\rm gate}$ or the pulse shape). As we will explain below, the difference
\begin{equation}
\theta \equiv \xi_2 - \xi_1 =  - \int_{0}^{t_{\rm gate}} \! \! \epsilon_{1} \, dt 
\label{theta definition}
\end{equation}
determines whether or not the ancilla becomes paralyzed during a leakage event. Note that $\theta$ can be varied during an experiment by changing the gate time. 

\subsection{Ancilla-assisted measurement with decoherence}
  
As shown in Fig.~\ref{fig:circuit1}, each measurement cycle consists of ancilla initialization, three gate operations, and ancilla readout.  Assuming ideal gates, the data qutrit after the first cycle is projected to a computational $\ket{0}$ or $\ket{1}$ state depending on the observed state of the ancilla [recall (\ref{ideal final state})]. In the absence of any errors, the measurement outcome of the ancilla remains unaltered thereafter. However, the situation is different in the presence of decoherence.

In order to model the effects of decoherence on the measurement outcomes of the ancilla, we assume that the readout and reset operations are instantaneous, while the Hadamard and {\sf CZ} gates take 10 and $25 \, {\rm ns}$ respectively. We also assume that amplitude damping is the only source of decoherence, in which case the single-qutrit Kraus matrices can be written as 
\BEq
\label{eq:qutritKrausAD}
\begin{array}{l}
E_1=\left[\begin{array}{ccc}
 1 & 0 & 0 \\
 0 & \sqrt{1-\lambda_1} & 0 \\
 0 & 0 & \sqrt{1-\lambda_2} \\
 \end{array}\right], \\ \\
 E_2=\left[\begin{array}{ccc}
 0 & \sqrt{\lambda_1} & 0 \\
 0 & 0 & 0 \\
 0 & 0 & 0 \\
 \end{array}\right], \\ \\
 E_{\rm 3}=\left[\begin{array}{ccc}
 0 & 0 & \sqrt{\lambda_2} \\
 0 & 0 & 0 \\
 0 & 0 & 0 \\
 \end{array}\right].
 \end{array}
\EEq 
For an operation of time duration ${\Delta}t$,
\BEq
\lambda_m = 1-e^{-m \, {\Delta}t /T_1}.
\EEq

\begin{figure*}[htb]
\includegraphics[angle=0,width=\textwidth]{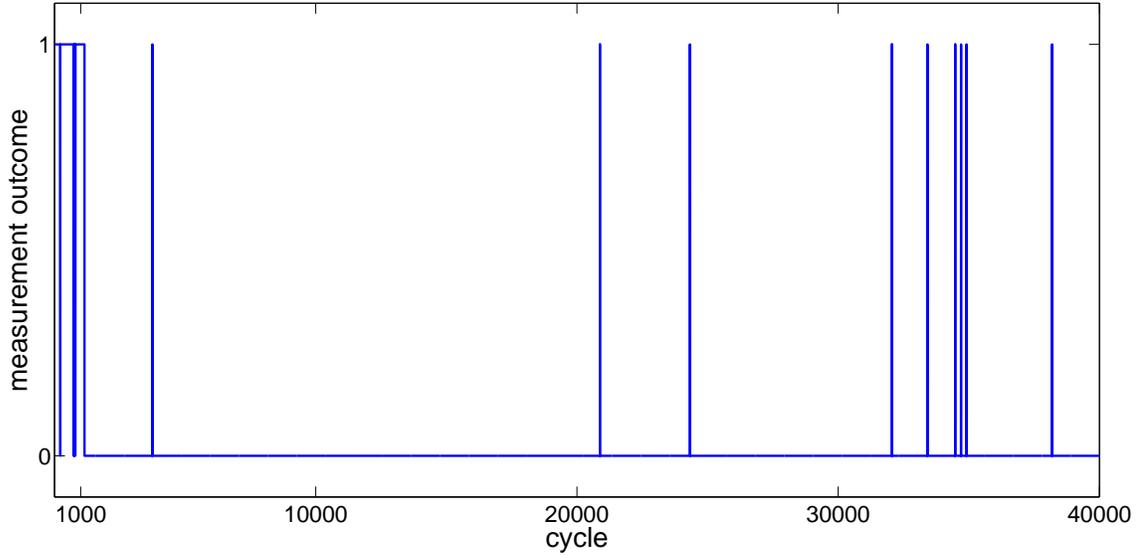}
\caption{(Color online) Simulated repeated readout of the ancilla qutrit in the presence of amplitude damping. Single peaks, upward or downward, indicate errors on the ancilla. Data errors result in steps; an example is shown near cycle 1000. In this simulation we assume $T_1=40 \, \mu{\rm s}$, $T_2 = 2 T_1$, and $t_{\rm cycle}=45 \, {\rm ns}$.}
\label{fig:circuit1Decoherence}
\end{figure*}

We simulate the ancilla-assisted measurement protocol for an ideal {\sf CZ} gate but in the presence of decoherence, for 40,000 consecutive cycles, and Fig.~\ref{fig:circuit1Decoherence} shows a typical outcome. The duration $t_{\rm cycle}$ of each complete measurement cycle is $45 \, {\rm ns}$ (one {\sf CZ} gate plus two Hadamards). Initially, the data qutrit is in state $\ket{1}$, and a single downward peak denotes an error on the ancilla. Near the $1000^{\rm th}$ cycle the data qutrit relaxes to $\ket{0}$ due to decoherence, and once in the ground state it stays there forever. The remaining upward peaks are caused by decoherence on the ancilla qutrit. Since the ancilla gets reset at the end of every cycle, such errors are manifested as single peaks. Note that if the initial state of the two-qutrit system is inside the computational subspace, it does not leak to non-computational states and therefore Fig.~\ref{fig:circuit1Decoherence} is insensitive to the values of the $\xi_i$.
 
\section{Non-ideal {\sf CZ} gate}
\label{sec:NonIdealCZ}
 
In this section, we first discuss how a non-ideal {\sf CZ} gate is parameterized and then investigate its action on the ancilla-assisted qubit measurement.

\subsection{Parameterization of the non-ideal CZ gate}
\label{sec:parameterizationNICZ}

Let us  first give a brief review of the dominant intrinsic error mechanisms that are relevant for the Strauch {\sf CZ} gate \cite{PhysRevLett.91.167005,PhysRevA.87.022309}; the Hadamards are always assumed to be ideal [see (\ref{H definition})]. The {\sf CZ} gate of Strauch {\it et al.}~\cite{PhysRevLett.91.167005} is performed by using the avoided level crossing between the $\ket{11}$ and $\ket{20}$ states at $\epsilon_1=\epsilon_2+\eta_1$. Although the other states are detuned from each other at this anticrossing point, a small amount of nonadiabatic population transfer is unavoidable, and these nonadiabatic excitations dominate the intrinsic gate errors \cite{PhysRevA.87.022309}. These errors can be thought of as producing a second unitary matrix whose generator $S^\prime$ can be parameterized, in the basis (\ref{two qutrit basis}), as
\BW
\BEq
\label{eq:SNI}
{ 
S^\prime=\left(\begin{array}{ccccccccc}
0 & 0 & 0 & 0 & 0 & 0 & 0 & 0 & 0 \\
0 & \zeta_1 & 0 & i\chi_{1}e^{i\phi_1} & 0 & 0 & 0 & 0 & 0 \\
0 & 0 &  0 & 0 & i\chi_{2}e^{i\phi_2} & 0 & 0 & 0 & 0 \\
0 & -i\chi_{1}e^{-i\phi_1} & 0 & \zeta_2 & 0 & 0 & 0 & 0 & 0 \\
0 & 0 & -i\chi_{2}e^{-i\phi_2} & 0 & \zeta_3 & 0 &  i\chi_{3}e^{i\phi_3} & 0 & 0 \\
0 & 0 & 0 & 0 & 0 & 0 & 0 & i\chi_{4}e^{i\phi_4} & 0 \\
0 & 0 & 0 & 0 & -i\chi_{3}e^{-i\phi_3} & 0 & \zeta_4 & 0 & 0 \\
0 & 0 & 0 & 0 & 0 & -i\chi_{4}e^{-i\phi_4}  & 0 & 0 & 0 \\
0 & 0 & 0 & 0 & 0 & 0 & 0 & 0 & 0 \\
\end{array}\right)}. \ \ 
\EEq
\EW
The complete non-ideal {\sf CZ} gate is
\BEq
\label{eq:UCZprime}
U_{\rm CZ}=e^{iS^\prime}e^{iS} \approx e^{i(S+S^\prime)},
\EEq
where $S$ is the generator (\ref{eq:SCZ}) of the ideal {\sf CZ} gate. The parameters $\chi_i$ and $\zeta_i$ in (\ref{eq:SNI}) are small, while the angles $\phi_i$ take arbitrary values between $0$ and $2\pi$. $\zeta_1$ and $\zeta_2$ parameterize the errors occurring during pre and post $\sigma^z$ rotations, and $\zeta_3$ and $\zeta_4$ denote the controlled-phase error for the $\ket{11}$ and $\ket{20}$ channels. In our simulations we assume $\chi_i=\zeta_i=10^{-2}$ for all $i=1, \ldots , 4$. Because population transfer probability scales with ${|\chi_i|}^{2}$, our choice of parameters bounds the intrinsic gate errors to about $10^{-4}$.

\subsection{Leakage events and ancilla paralysis}
\label{sec:LeakageSignature} 

The {\sf CZ} gate (\ref{eq:UCZprime}) produces, on any $|11\rangle$ input component, a small amplitude of $|02\rangle$ (the amount determined by $\chi_2$) and $|20\rangle$ (determined by $\chi_3$). 
A $|20\rangle$ component either results in the possibility of an ancilla readout of $|2\rangle$---if the readout protocol distinguishes $|1\rangle$ and $|2\rangle$---or the possiblity of an isolated ancilla error if it does not. Neither case compromises fault-tolerance. The parameter $\chi_2$ is responsible for data qubit leakage events. By a leakage {\it event} we mean a near-unity population of the data $|2\rangle$
state.

The principal mechanism producing a leakage event is the abrupt, nonlinear transformation on the data qutrit induced by the ancilla measurement. We denote these transformations by ${\bf T}_0$, ${\bf T}_1$, and ${\bf T}_2$, where the subscript corresponds to the ancilla readout result. Repeatedly measuring the ancilla applies a random sequence of the ${\bf T}$ maps to the data qutrit.

For the model, gate implementation, and parameter values considered in this work, the map ${\bf T}_0$ is primarily responsible for the observed leakage events. Although the general form of ${\bf T}_0$ is quite complex, it is possible to construct a simple special case of it that exhibits the essential features. To do this we choose simplified parameter values
\begin{eqnarray}
\xi_1 &=& \pi, \nonumber \\
\phi_i &=& 0, \nonumber \\
\zeta_i &=& 0, \nonumber \\
\chi_3 &=& 0, \nonumber \\
\chi_4 &=& 0 ,
\end{eqnarray}
and calculate the action of the non-ideal measurement circuit on an arbitrary data qutrit state (up to but not including readout)
\begin{equation}
|\psi_{\rm D}\rangle = a |0\rangle + b |1\rangle + c |2\rangle .
\label{general data qutrit state}
\end{equation}
We find (in the $|{\rm AD}\rangle$ basis) that
\BW
\begin{eqnarray}
a |00\rangle + b |01\rangle + c |02\rangle &\rightarrow& 
|0\rangle \otimes \bigg[ 
\bigg( \frac{a}{2} +  \frac{a}{2}  \cos \chi_1 + \frac{b}{2}  \sin \chi_1
\bigg)  |0\rangle + \bigg( \frac{b}{2} \cos \chi_1  -  \frac{a}{2}  \sin \chi_1 - \frac{b}{2}  \cos \chi_2 - \frac{c}{2}  \sin \chi_2 \bigg)  |1\rangle \nonumber \\
&+& \bigg( \frac{c}{2} +  \frac{b}{2}  \sin \chi_2 - \frac{c}{2}  \cos \chi_2
\bigg)  |2\rangle \bigg] + |1\rangle \otimes \bigg[ 
\bigg( \frac{a}{2} -  \frac{a}{2}  \cos \chi_1 - \frac{b}{2}  \sin \chi_1
\bigg)  |0\rangle \nonumber \\
&+& \bigg( \frac{b}{2} \cos \chi_1 - \frac{a}{2}  \sin \chi_1 + \frac{b}{2}  \cos \chi_2 + \frac{c}{2}  \sin \chi_2 \bigg)  |1\rangle - \bigg( \frac{c}{2} -  \frac{b}{2}  \sin \chi_2 + \frac{c}{2}  \cos \chi_2
\bigg)  |2\rangle \bigg].
\end{eqnarray}
\EW
An ancilla readout result of $|0\rangle$ then induces the map ${\bf T}_0$ given by
\begin{eqnarray}
a \rightarrow a' &=& \frac{a+  a  \cos \chi_1 + b \sin \chi_1}{\sqrt{{\cal N}}} ,  \nonumber \\
b \rightarrow b' &=& \frac{b \cos \chi_1  -  a  \sin \chi_1 - b \cos \chi_2 - c  \sin \chi_2}{\sqrt{{\cal N}}} , \nonumber \\
c \rightarrow c' &=& \frac{c + b \sin \chi_2 
- c  \cos \chi_2}{\sqrt{{\cal N}}} ,
\label{T0 map}
\end{eqnarray}
where
\begin{eqnarray}
{\cal N} &\equiv& \big|a+  a  \cos \chi_1 + b \sin \chi_1 \big|^2 \nonumber \\
&+& \big|b \cos \chi_1  -  a  \sin \chi_1 - b \cos \chi_2 - c  \sin \chi_2 \big|^2 \nonumber \\
&+& \big|c + b \sin \chi_2 - c  \cos \chi_2 \big|^2.
\end{eqnarray}
Using (\ref{T0 map}) we find that in the limit $\chi_1 =  0$ and $\chi_2 \rightarrow 0$ the data qutrit prepared in the $|1\rangle$ state transforms as
\begin{equation}
{\bf T}_0 \,  |1\rangle = |2\rangle .
\label{leakage map}
\end{equation}
Our simulations confirm that the dominant mechanism for producing a leakage event is the process (\ref{leakage map}). Physically, the reason why the ${\bf T}_0$ map plays the dominant role for data qubit leakage is the following: The data qubit leakage events occur in the double-excitation subspace spanned by $\{\ket{02}, \ket{11}, \ket{20}\}$. Therefore, in order for the data qubit to be in the $\ket{2}$ state, the ancilla must be in the $\ket{0}$ state so that the entire two qubit system remains in the double excitation subspace.

\begin{figure*}[htb]
\includegraphics[angle=0,width=\textwidth]{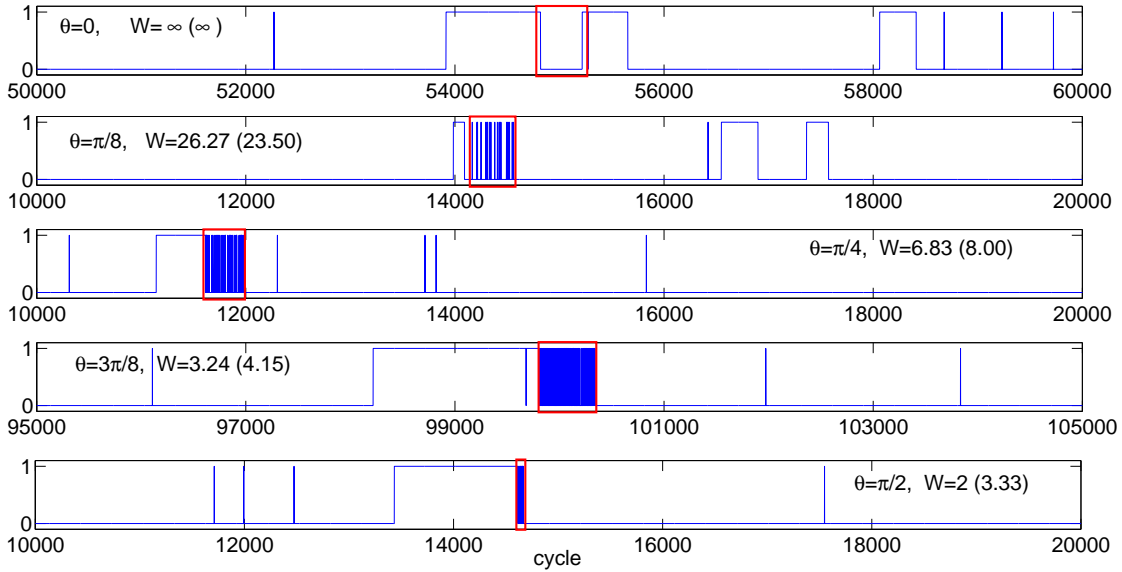}
\caption{(Color online) Simulated sequential measurements of the ancilla qubit. The readout values $|0\rangle$ or $|1\rangle$ are shown as a function of measurement cycle number. Red rectangles signify leakage events, where the data $|2\rangle$ state probability is close to unity.  Random ancilla oscillations during the leakage events are observed except when $\theta \approx 0$. Two values of $W$ are given for each trace: the theoretical value from (\ref{W estimate}) and a value, shown in parentheses, numerically computed from the simulation. The simulations assume $T_1=40 \, {\rm \mu s}$, $T_2=2T_1$, $\chi_i=\zeta_i=10^{-2}$ for all $i=1, \ldots , 4$, and random values of phase angle parameters consistent with the indicated values of $\theta$.}
\label{fig:ancillaVsXi3}
\end{figure*}

Once leaked, the data qutrit remains in the $|2\rangle$ state (for many cycles) until it either undergoes a nonadiabatic ``reverse-leakage" transition or it relaxes back  to the computational subspace.
The behaviour of the ancilla during a leakage event depends on the values of $\xi_1$ and $\xi_2$ in (\ref{eq:SCZ}). While the data qubit is in the $\ket{2}$ state, the two-qutrit system is restricted to the subspace spanned by
\begin{equation} 
\big\lbrace \ket{02},\ket{12} \big\rbrace,
\label{leakage subspace}
\end{equation}
because the $|22\rangle$ state is decoupled and remains unoccupied. In this subspace, the {\sf CZ} gate (\ref{eq:UCZprime}) acts as
\begin{equation}
\exp \bigg[ i
{\left(\begin{array}{cc} \xi_1 & 0  \\ 0 & \xi_2  \\
\end{array}\right)} \bigg],
\end{equation}
and therefore performs a $z$ rotation on the ancilla by an angle (\ref{theta definition}). The Hadamards in Fig.~\ref{fig:circuit1} convert this to an $x$ rotation [see (\ref{theta definition})]
\begin{equation}
e^{-i(\theta/2)\sigma^x}
\end{equation}
acting on the initial ancilla state $|0\rangle$. Therefore, during a leakage event, while the data qubit is locked in the $|2\rangle$ state, the state of the ancilla after every cycle is
\begin{equation}
\cos \textstyle{\frac{\theta}{2}} \, |0\rangle + \sin \textstyle{\frac{\theta}{2}} \, |1\rangle,
\label{final ancilla state during leakage}
\end{equation}
and upon measurement the ancilla qubit reads $|0\rangle$ with probability $\cos^2(\theta/2)$. 

For example, if 
\begin{equation}
\theta \ {\rm mod} \, \pi =\frac{\pi}{2},
\end{equation}
we will observe random ancilla outcomes with equal probabilities for observing $|0\rangle$ and $|1\rangle$. This type of leakage event is simple to detect (and possibly correct). However, if
\begin{equation}
\theta \ {\rm mod} \, \pi = 0,
\end{equation}
then the ancilla will always read $|0\rangle$, cycle after cycle, giving no indication of the data error and thereby compromising fault-tolerance. We refer to this dangerous phenomena as ancilla {\it paralysis}.

Figure \ref{fig:ancillaVsXi3} shows the readout values generated from the sequential measurements of the ancilla qubit for different choices of $\theta$, including all error process contained in the non-ideal {\sf CZ} gate (\ref{eq:UCZprime}) as well as decoherence. While we observe random oscillations for larger values of $\theta$, no such signature is present for $\theta = 0$. In order to quantify the paralysis of the ancilla we define a metric $W$, which is the average spacing---number of 
cycles---between consecutive readouts of $|1\rangle$. In the absence of decoherence, we can estimate it [see (\ref{final ancilla state during leakage})] as
\begin{equation}
W = \csc^2 (\theta/2),
\label{W estimate}
\end{equation}
which agrees well with the numerical simulations.

The detectability of a leakage event depends on whether $W$ is small enough to be observed in the presence of a background value $W^*$ resulting from decoherence (and possibly other errors). For example, in the simulations of Fig.~\ref{fig:ancillaVsXi3}, which have $T_1=40$ $\mu{s}$ and $T_2=2T_1$, the average spacing between ancilla $|1\rangle$ peaks away from the leakage events is 2381 cycles, which is not too far from the crude theoretical estimate
\begin{equation}
W^* \approx \frac{2 T_1}{t_{\rm cycle}} = 1778,
\label{W* estimate}
\end{equation}
using $t_{\rm cycle} = 45 \, {\rm ns}$. The estimate in
(\ref{W* estimate}) can be derived from the Pauli twirling approximation for qubit decoherence \cite{PhysRevA.86.062318,Geller&ZhouPRA13}, which predicts $\sigma^x$ and $\sigma^y$ errors on the ancilla with probability $p_{\scriptscriptstyle X} = p_{\scriptscriptstyle Y} = t_{\rm cycle}/4 T_1$, leading to a total bit-flip probability $p_{\scriptscriptstyle X}+p_{\scriptscriptstyle Y}$ of $t_{\rm cycle}/2 T_1$. We can then use (\ref{W estimate}) to estimate the critical value of $\theta$ separating the region of dangerous ancilla paralysis and that of ordinary leakage, namely
\begin{equation}
\theta^* = 2 \, {\rm csc}^{-1} \sqrt{W^*} \approx 2 \, {\rm csc}^{-1} \bigg( \sqrt{\frac{2 T_1}{t_{\rm cycle}} } \bigg),
\end{equation}
which is $\theta^* = 0.04$ in the simulations reported here. {\sf CZ} gates with $\theta \ {\rm mod} \, \pi < \theta^*$ are susceptible to undetectable leakage events. 

 \section{Conclusions}
 \label{sec:Conclusion}
 
We have investigated the physics of leakage within a two-qubit superconducting quantum error detection circuit, where the error detection is performed via repeated ancilla-assisted measurement of the $\sigma^z$ operator for a data qubit. We have observed that leakage does not propagate to a neighboring qubit via the two-qubit entangling operations, a necessary condition for gate-based  quantum error correction to be possible. However, interacting with a leaked data qubit can either randomize or paralyze a measurement qubit, both of which will be difficult to handle with existing error correction techniques, especially when long-lived leakage events lead to long strings of time-correlated measurement errors.

Our results are clearly relevant for topological error correction protocols, such as surface or toric code, where the measurements of stabilizer operators are performed via similar ancilla-assisted schemes. In standard surface code error correction, physical qubits are arranged in a 2D square lattice and a sequence of gate operations is performed in order to measure the 3- and 4-qubit Pauli operators, and repeated for many cycles. Each error correction cycle, therefore, comprises $4$ {\sf CNOT} (or {\sf CZ}) operations between a syndrome qubit and its $4$ nearest-neighbor data qubits. Since data qubits are never measured, a single leakage event on a data qubit during any of the $4$ {\sf CNOT} (or {\sf CZ}) gates would destroy the entire multi-qubit operator measurement by generating noisy outcomes in the ancilla qubit (or irrelevant ancilla outputs in case of paralysis). In our work, we have analyzed a single qubit operator measurement and it is important to emphasize that the effect of such leakage events could be enormous for a long-running quantum algorithm in a million-qubit quantum computer.

One approach to suppress the effects of leakage is to use a stabilizer-based topological error-correcting code for qudits, a direction where an active theoretical research is currently underway \cite{1751-8121-40-13-013,PhysRevA.78.062315,2011arXiv1101.1519G,tagkey19991749,PhysRevA.71.042315,Jafarizadeh2008EPJD,PhysRevA.84.052306}. Another approach to cope with leakage is to supplement the standard topological error-correcting techniques with additional steps, where each qubit is repeatedly initialized, operated on by gates, and measured, thereby systematically removing leakage errors from all qubits in the array, at the cost of some additional operations. Such additional operations involve a quantum circuitry to teleport the quantum information from the data qubits to ancillas for a given cycle and use the ancillas as data qubits for the consecutive cycle. Such a software approach has recently been adopted \cite{2013arXiv1308.6642F} by one of the authors based on the work presented here, where the cost of such additional operations and enhancement of performance against leakage errors are quantified in the context of a repetition code.

\begin{acknowledgments}
This research was funded by the US Office of the Director of National Intelligence (ODNI), Intelligence Advanced Research Projects Activity (IARPA), through the US Army Research Office grant No.~W911NF-10-1-0334. All statements of fact, opinion or conclusions contained herein are those of the authors and should not be construed as representing the official views or policies of IARPA, the ODNI, or the US Government. J.G. gratefully acknowledges the financial support from NSERC, AITF and University of Calgary's Eyes High Fellowship Program. Part of this work was carried out while M.G.~was a Lady Davis Visiting Professor in the Racah Institute of Physics at Hebrew University, Jerusalem.
\end{acknowledgments}

\bibliography{leakage}

%merlin.mbs apsrev4-1.bst 2010-07-25 4.21a (PWD, AO, DPC) hacked
%Control: key (0)
%Control: author (8) initials jnrlst
%Control: editor formatted (1) identically to author
%Control: production of article title (-1) disabled
%Control: page (0) single
%Control: year (1) truncated
%Control: production of eprint (0) enabled
\begin{thebibliography}{23}%
\makeatletter
\providecommand \@ifxundefined [1]{%
 \@ifx{#1\undefined}
}%
\providecommand \@ifnum [1]{%
 \ifnum #1\expandafter \@firstoftwo
 \else \expandafter \@secondoftwo
 \fi
}%
\providecommand \@ifx [1]{%
 \ifx #1\expandafter \@firstoftwo
 \else \expandafter \@secondoftwo
 \fi
}%
\providecommand \natexlab [1]{#1}%
\providecommand \enquote  [1]{``#1''}%
\providecommand \bibnamefont  [1]{#1}%
\providecommand \bibfnamefont [1]{#1}%
\providecommand \citenamefont [1]{#1}%
\providecommand \href@noop [0]{\@secondoftwo}%
\providecommand \href [0]{\begingroup \@sanitize@url \@href}%
\providecommand \@href[1]{\@@startlink{#1}\@@href}%
\providecommand \@@href[1]{\endgroup#1\@@endlink}%
\providecommand \@sanitize@url [0]{\catcode `\\12\catcode `\$12\catcode
  `\&12\catcode `\#12\catcode `\^12\catcode `\_12\catcode `\%12\relax}%
\providecommand \@@startlink[1]{}%
\providecommand \@@endlink[0]{}%
\providecommand \url  [0]{\begingroup\@sanitize@url \@url }%
\providecommand \@url [1]{\endgroup\@href {#1}{\urlprefix }}%
\providecommand \urlprefix  [0]{URL }%
\providecommand \Eprint [0]{\href }%
\providecommand \doibase [0]{http://dx.doi.org/}%
\providecommand \selectlanguage [0]{\@gobble}%
\providecommand \bibinfo  [0]{\@secondoftwo}%
\providecommand \bibfield  [0]{\@secondoftwo}%
\providecommand \translation [1]{[#1]}%
\providecommand \BibitemOpen [0]{}%
\providecommand \bibitemStop [0]{}%
\providecommand \bibitemNoStop [0]{.\EOS\space}%
\providecommand \EOS [0]{\spacefactor3000\relax}%
\providecommand \BibitemShut  [1]{\csname bibitem#1\endcsname}%
\let\auto@bib@innerbib\@empty
%</preamble>
\bibitem [{\citenamefont {Fazio}\ \emph {et~al.}(1999)\citenamefont {Fazio},
  \citenamefont {Palma},\ and\ \citenamefont {Siewert}}]{PhysRevLett.83.5385}%
  \BibitemOpen
  \bibfield  {author} {\bibinfo {author} {\bibfnamefont {R.}~\bibnamefont
  {Fazio}}, \bibinfo {author} {\bibfnamefont {G.~M.}\ \bibnamefont {Palma}}, \
  and\ \bibinfo {author} {\bibfnamefont {J.}~\bibnamefont {Siewert}},\ }\href
  {\doibase 10.1103/PhysRevLett.83.5385} {\bibfield  {journal} {\bibinfo
  {journal} {Phys. Rev. Lett.}\ }\textbf {\bibinfo {volume} {83}},\ \bibinfo
  {pages} {5385} (\bibinfo {year} {1999})}\BibitemShut {NoStop}%
\bibitem [{\citenamefont {Ferr\'on}\ and\ \citenamefont
  {Dom\'inguez}(2010)}]{PhysRevB.81.104505}%
  \BibitemOpen
  \bibfield  {author} {\bibinfo {author} {\bibfnamefont {A.}~\bibnamefont
  {Ferr\'on}}\ and\ \bibinfo {author} {\bibfnamefont {D.}~\bibnamefont
  {Dom\'inguez}},\ }\href {\doibase 10.1103/PhysRevB.81.104505} {\bibfield
  {journal} {\bibinfo  {journal} {Phys. Rev. B}\ }\textbf {\bibinfo {volume}
  {81}},\ \bibinfo {pages} {104505} (\bibinfo {year} {2010})}\BibitemShut
  {NoStop}%
\bibitem [{\citenamefont {Ainsworth}\ and\ \citenamefont
  {Slingerland}(2011)}]{1367-2630-13-6-065030}%
  \BibitemOpen
  \bibfield  {author} {\bibinfo {author} {\bibfnamefont {R.}~\bibnamefont
  {Ainsworth}}\ and\ \bibinfo {author} {\bibfnamefont {J.~K.}\ \bibnamefont
  {Slingerland}},\ }\href {http://stacks.iop.org/1367-2630/13/i=6/a=065030}
  {\bibfield  {journal} {\bibinfo  {journal} {New Journal of Physics}\ }\textbf
  {\bibinfo {volume} {13}},\ \bibinfo {pages} {065030} (\bibinfo {year}
  {2011})}\BibitemShut {NoStop}%
\bibitem [{\citenamefont {You}\ and\ \citenamefont {Nori}(2005)}]{You2005}%
  \BibitemOpen
  \bibfield  {author} {\bibinfo {author} {\bibfnamefont {J.~Q.}\ \bibnamefont
  {You}}\ and\ \bibinfo {author} {\bibfnamefont {F.}~\bibnamefont {Nori}},\
  }\href@noop {} {\bibfield  {journal} {\bibinfo  {journal} {Physics Today}\
  }\textbf {\bibinfo {volume} {58}},\ \bibinfo {pages} {42} (\bibinfo {year}
  {2005})},\ \bibinfo {note} {p. 42}\BibitemShut {NoStop}%
\bibitem [{\citenamefont {Neeley}\ \emph {et~al.}(2009)\citenamefont {Neeley},
  \citenamefont {Ansmann}, \citenamefont {Bialczak}, \citenamefont {Hofheinz},
  \citenamefont {Lucero}, \citenamefont {O'Connell}, \citenamefont {Sank},
  \citenamefont {Wang}, \citenamefont {Wenner}, \citenamefont {Cleland},
  \citenamefont {Geller},\ and\ \citenamefont {Martinis}}]{Neeley07082009}%
  \BibitemOpen
  \bibfield  {author} {\bibinfo {author} {\bibfnamefont {M.}~\bibnamefont
  {Neeley}}, \bibinfo {author} {\bibfnamefont {M.}~\bibnamefont {Ansmann}},
  \bibinfo {author} {\bibfnamefont {R.~C.}\ \bibnamefont {Bialczak}}, \bibinfo
  {author} {\bibfnamefont {M.}~\bibnamefont {Hofheinz}}, \bibinfo {author}
  {\bibfnamefont {E.}~\bibnamefont {Lucero}}, \bibinfo {author} {\bibfnamefont
  {A.~D.}\ \bibnamefont {O'Connell}}, \bibinfo {author} {\bibfnamefont
  {D.}~\bibnamefont {Sank}}, \bibinfo {author} {\bibfnamefont {H.}~\bibnamefont
  {Wang}}, \bibinfo {author} {\bibfnamefont {J.}~\bibnamefont {Wenner}},
  \bibinfo {author} {\bibfnamefont {A.~N.}\ \bibnamefont {Cleland}}, \bibinfo
  {author} {\bibfnamefont {M.~R.}\ \bibnamefont {Geller}}, \ and\ \bibinfo
  {author} {\bibfnamefont {J.~M.}\ \bibnamefont {Martinis}},\ }\href {\doibase
  10.1126/science.1173440} {\bibfield  {journal} {\bibinfo  {journal}
  {Science}\ }\textbf {\bibinfo {volume} {325}},\ \bibinfo {pages} {722}
  (\bibinfo {year} {2009})}\BibitemShut {NoStop}%
\bibitem [{\citenamefont {Strauch}\ \emph {et~al.}(2003)\citenamefont
  {Strauch}, \citenamefont {Johnson}, \citenamefont {Dragt}, \citenamefont
  {Lobb}, \citenamefont {Anderson},\ and\ \citenamefont
  {Wellstood}}]{PhysRevLett.91.167005}%
  \BibitemOpen
  \bibfield  {author} {\bibinfo {author} {\bibfnamefont {F.~W.}\ \bibnamefont
  {Strauch}}, \bibinfo {author} {\bibfnamefont {P.~R.}\ \bibnamefont
  {Johnson}}, \bibinfo {author} {\bibfnamefont {A.~J.}\ \bibnamefont {Dragt}},
  \bibinfo {author} {\bibfnamefont {C.~J.}\ \bibnamefont {Lobb}}, \bibinfo
  {author} {\bibfnamefont {J.~R.}\ \bibnamefont {Anderson}}, \ and\ \bibinfo
  {author} {\bibfnamefont {F.~C.}\ \bibnamefont {Wellstood}},\ }\href {\doibase
  10.1103/PhysRevLett.91.167005} {\bibfield  {journal} {\bibinfo  {journal}
  {Phys. Rev. Lett.}\ }\textbf {\bibinfo {volume} {91}},\ \bibinfo {pages}
  {167005} (\bibinfo {year} {2003})}\BibitemShut {NoStop}%
\bibitem [{\citenamefont {DiCarlo}\ \emph {et~al.}(2009)\citenamefont
  {DiCarlo}, \citenamefont {Chow}, \citenamefont {Gambetta}, \citenamefont
  {Bishop}, \citenamefont {Johnson}, \citenamefont {Schuster}, \citenamefont
  {Majer}, \citenamefont {Blais}, \citenamefont {Frunzio}, \citenamefont
  {Girvin},\ and\ \citenamefont {Schoelkopf}}]{DiCarlo2009}%
  \BibitemOpen
  \bibfield  {author} {\bibinfo {author} {\bibfnamefont {L.}~\bibnamefont
  {DiCarlo}}, \bibinfo {author} {\bibfnamefont {J.~M.}\ \bibnamefont {Chow}},
  \bibinfo {author} {\bibfnamefont {J.~M.}\ \bibnamefont {Gambetta}}, \bibinfo
  {author} {\bibfnamefont {L.~S.}\ \bibnamefont {Bishop}}, \bibinfo {author}
  {\bibfnamefont {B.~R.}\ \bibnamefont {Johnson}}, \bibinfo {author}
  {\bibfnamefont {D.~I.}\ \bibnamefont {Schuster}}, \bibinfo {author}
  {\bibfnamefont {J.}~\bibnamefont {Majer}}, \bibinfo {author} {\bibfnamefont
  {A.}~\bibnamefont {Blais}}, \bibinfo {author} {\bibfnamefont
  {L.}~\bibnamefont {Frunzio}}, \bibinfo {author} {\bibfnamefont {S.~M.}\
  \bibnamefont {Girvin}}, \ and\ \bibinfo {author} {\bibfnamefont {R.~J.}\
  \bibnamefont {Schoelkopf}},\ }\href {http://dx.doi.org/10.1038/nature08121}
  {\bibfield  {journal} {\bibinfo  {journal} {Nature}\ }\textbf {\bibinfo
  {volume} {460}},\ \bibinfo {pages} {240} (\bibinfo {year}
  {2009})}\BibitemShut {NoStop}%
\bibitem [{\citenamefont {Ghosh}\ \emph {et~al.}(2013)\citenamefont {Ghosh},
  \citenamefont {Galiautdinov}, \citenamefont {Zhou}, \citenamefont {Korotkov},
  \citenamefont {Martinis},\ and\ \citenamefont {Geller}}]{PhysRevA.87.022309}%
  \BibitemOpen
  \bibfield  {author} {\bibinfo {author} {\bibfnamefont {J.}~\bibnamefont
  {Ghosh}}, \bibinfo {author} {\bibfnamefont {A.}~\bibnamefont {Galiautdinov}},
  \bibinfo {author} {\bibfnamefont {Z.}~\bibnamefont {Zhou}}, \bibinfo {author}
  {\bibfnamefont {A.~N.}\ \bibnamefont {Korotkov}}, \bibinfo {author}
  {\bibfnamefont {J.~M.}\ \bibnamefont {Martinis}}, \ and\ \bibinfo {author}
  {\bibfnamefont {M.~R.}\ \bibnamefont {Geller}},\ }\href {\doibase
  10.1103/PhysRevA.87.022309} {\bibfield  {journal} {\bibinfo  {journal} {Phys.
  Rev. A}\ }\textbf {\bibinfo {volume} {87}},\ \bibinfo {pages} {022309}
  (\bibinfo {year} {2013})}\BibitemShut {NoStop}%
\bibitem [{\citenamefont {Motzoi}\ \emph {et~al.}(2009)\citenamefont {Motzoi},
  \citenamefont {Gambetta}, \citenamefont {Rebentrost},\ and\ \citenamefont
  {Wilhelm}}]{Motzoi2009}%
  \BibitemOpen
  \bibfield  {author} {\bibinfo {author} {\bibfnamefont {F.}~\bibnamefont
  {Motzoi}}, \bibinfo {author} {\bibfnamefont {J.~M.}\ \bibnamefont
  {Gambetta}}, \bibinfo {author} {\bibfnamefont {P.}~\bibnamefont
  {Rebentrost}}, \ and\ \bibinfo {author} {\bibfnamefont {F.~K.}\ \bibnamefont
  {Wilhelm}},\ }\href {\doibase 10.1103/PhysRevLett.103.110501} {\bibfield
  {journal} {\bibinfo  {journal} {Phys. Rev. Lett.}\ }\textbf {\bibinfo
  {volume} {103}},\ \bibinfo {pages} {110501} (\bibinfo {year}
  {2009})}\BibitemShut {NoStop}%
\bibitem [{\citenamefont {Egger}\ and\ \citenamefont
  {Wilhelm}(2014)}]{2013arXiv1306.6894E}%
  \BibitemOpen
  \bibfield  {author} {\bibinfo {author} {\bibfnamefont {D.~J.}\ \bibnamefont
  {Egger}}\ and\ \bibinfo {author} {\bibfnamefont {F.~K.}\ \bibnamefont
  {Wilhelm}},\ }\href {http://stacks.iop.org/0953-2048/27/i=1/a=014001}
  {\bibfield  {journal} {\bibinfo  {journal} {Superconductor Science and
  Technology}\ }\textbf {\bibinfo {volume} {27}},\ \bibinfo {pages} {014001}
  (\bibinfo {year} {2014})}\BibitemShut {NoStop}%
\bibitem [{\citenamefont {Devitt}\ \emph {et~al.}(2013)\citenamefont {Devitt},
  \citenamefont {Munro},\ and\ \citenamefont {Nemoto}}]{0034-4885-76-7-076001}%
  \BibitemOpen
  \bibfield  {author} {\bibinfo {author} {\bibfnamefont {S.~J.}\ \bibnamefont
  {Devitt}}, \bibinfo {author} {\bibfnamefont {W.~J.}\ \bibnamefont {Munro}}, \
  and\ \bibinfo {author} {\bibfnamefont {K.}~\bibnamefont {Nemoto}},\ }\href
  {http://stacks.iop.org/0034-4885/76/i=7/a=076001} {\bibfield  {journal}
  {\bibinfo  {journal} {Reports on Progress in Physics}\ }\textbf {\bibinfo
  {volume} {76}},\ \bibinfo {pages} {076001} (\bibinfo {year}
  {2013})}\BibitemShut {NoStop}%
\bibitem [{\citenamefont {Ghosh}\ \emph {et~al.}(2012)\citenamefont {Ghosh},
  \citenamefont {Fowler},\ and\ \citenamefont {Geller}}]{PhysRevA.86.062318}%
  \BibitemOpen
  \bibfield  {author} {\bibinfo {author} {\bibfnamefont {J.}~\bibnamefont
  {Ghosh}}, \bibinfo {author} {\bibfnamefont {A.~G.}\ \bibnamefont {Fowler}}, \
  and\ \bibinfo {author} {\bibfnamefont {M.~R.}\ \bibnamefont {Geller}},\
  }\href {\doibase 10.1103/PhysRevA.86.062318} {\bibfield  {journal} {\bibinfo
  {journal} {Phys. Rev. A}\ }\textbf {\bibinfo {volume} {86}},\ \bibinfo
  {pages} {062318} (\bibinfo {year} {2012})}\BibitemShut {NoStop}%
\bibitem [{\citenamefont {Fowler}\ \emph {et~al.}(2012)\citenamefont {Fowler},
  \citenamefont {Whiteside}, \citenamefont {McInnes},\ and\ \citenamefont
  {Rabbani}}]{PhysRevX.2.041003}%
  \BibitemOpen
  \bibfield  {author} {\bibinfo {author} {\bibfnamefont {A.~G.}\ \bibnamefont
  {Fowler}}, \bibinfo {author} {\bibfnamefont {A.~C.}\ \bibnamefont
  {Whiteside}}, \bibinfo {author} {\bibfnamefont {A.~L.}\ \bibnamefont
  {McInnes}}, \ and\ \bibinfo {author} {\bibfnamefont {A.}~\bibnamefont
  {Rabbani}},\ }\href {\doibase 10.1103/PhysRevX.2.041003} {\bibfield
  {journal} {\bibinfo  {journal} {Phys. Rev. X}\ }\textbf {\bibinfo {volume}
  {2}},\ \bibinfo {pages} {041003} (\bibinfo {year} {2012})}\BibitemShut
  {NoStop}%
\bibitem [{\citenamefont {Fowler}(2013)}]{2013arXiv1308.6642F}%
  \BibitemOpen
  \bibfield  {author} {\bibinfo {author} {\bibfnamefont {A.~G.}\ \bibnamefont
  {Fowler}},\ }\href {\doibase 10.1103/PhysRevA.88.042308} {\bibfield
  {journal} {\bibinfo  {journal} {Phys. Rev. A}\ }\textbf {\bibinfo {volume}
  {88}},\ \bibinfo {pages} {042308} (\bibinfo {year} {2013})}\BibitemShut
  {NoStop}%
\bibitem [{\citenamefont {Mariantoni}\ \emph {et~al.}(2011)\citenamefont
  {Mariantoni}, \citenamefont {Wang}, \citenamefont {Yamamoto}, \citenamefont
  {Neeley}, \citenamefont {Bialczak}, \citenamefont {Chen}, \citenamefont
  {Lenander}, \citenamefont {Lucero}, \citenamefont {O'Connell},
  \citenamefont {Sank}, \citenamefont {Weides}, \citenamefont {Wenner},
  \citenamefont {Yin}, \citenamefont {Zhao}, \citenamefont {Korotkov},
  \citenamefont {Cleland},\ and\ \citenamefont
  {Martinis}}]{Mariantoni07102011}%
  \BibitemOpen
  \bibfield  {author} {\bibinfo {author} {\bibfnamefont {M.}~\bibnamefont
  {Mariantoni}}, \bibinfo {author} {\bibfnamefont {H.}~\bibnamefont {Wang}},
  \bibinfo {author} {\bibfnamefont {T.}~\bibnamefont {Yamamoto}}, \bibinfo
  {author} {\bibfnamefont {M.}~\bibnamefont {Neeley}}, \bibinfo {author}
  {\bibfnamefont {R.~C.}\ \bibnamefont {Bialczak}}, \bibinfo {author}
  {\bibfnamefont {Y.}~\bibnamefont {Chen}}, \bibinfo {author} {\bibfnamefont
  {M.}~\bibnamefont {Lenander}}, \bibinfo {author} {\bibfnamefont
  {E.}~\bibnamefont {Lucero}}, \bibinfo {author} {\bibfnamefont {A.~D.}\
  \bibnamefont {O'Connell}}, \bibinfo {author} {\bibfnamefont
  {D.}~\bibnamefont {Sank}}, \bibinfo {author} {\bibfnamefont {M.}~\bibnamefont
  {Weides}}, \bibinfo {author} {\bibfnamefont {J.}~\bibnamefont {Wenner}},
  \bibinfo {author} {\bibfnamefont {Y.}~\bibnamefont {Yin}}, \bibinfo {author}
  {\bibfnamefont {J.}~\bibnamefont {Zhao}}, \bibinfo {author} {\bibfnamefont
  {A.~N.}\ \bibnamefont {Korotkov}}, \bibinfo {author} {\bibfnamefont {A.~N.}\
  \bibnamefont {Cleland}}, \ and\ \bibinfo {author} {\bibfnamefont {J.~M.}\
  \bibnamefont {Martinis}},\ }\href {\doibase 10.1126/science.1208517}
  {\bibfield  {journal} {\bibinfo  {journal} {Science}\ }\textbf {\bibinfo
  {volume} {334}},\ \bibinfo {pages} {61} (\bibinfo {year} {2011})}\BibitemShut
  {NoStop}%
\bibitem [{\citenamefont {Geller}\ and\ \citenamefont
  {Zhou}(2013)}]{Geller&ZhouPRA13}%
  \BibitemOpen
  \bibfield  {author} {\bibinfo {author} {\bibfnamefont {M.~R.}\ \bibnamefont
  {Geller}}\ and\ \bibinfo {author} {\bibfnamefont {Z.}~\bibnamefont {Zhou}},\
  }\href {\doibase 10.1103/PhysRevA.88.012314} {\bibfield  {journal} {\bibinfo
  {journal} {Phys. Rev. A}\ }\textbf {\bibinfo {volume} {88}},\ \bibinfo
  {pages} {012314} (\bibinfo {year} {2013})}\BibitemShut {NoStop}%
\bibitem [{\citenamefont {Bullock}\ and\ \citenamefont
  {Brennen}(2007)}]{1751-8121-40-13-013}%
  \BibitemOpen
  \bibfield  {author} {\bibinfo {author} {\bibfnamefont {S.~S.}\ \bibnamefont
  {Bullock}}\ and\ \bibinfo {author} {\bibfnamefont {G.~K.}\ \bibnamefont
  {Brennen}},\ }\href {\doibase 10.1088/1751-8113/40/13/013} {\bibfield
  {journal} {\bibinfo  {journal} {Journal of Physics A: Mathematical and
  Theoretical}\ }\textbf {\bibinfo {volume} {40}},\ \bibinfo {pages} {3481}
  (\bibinfo {year} {2007})}\BibitemShut {NoStop}%
\bibitem [{\citenamefont {Chen}\ \emph {et~al.}(2008)\citenamefont {Chen},
  \citenamefont {Zeng},\ and\ \citenamefont {Chuang}}]{PhysRevA.78.062315}%
  \BibitemOpen
  \bibfield  {author} {\bibinfo {author} {\bibfnamefont {X.}~\bibnamefont
  {Chen}}, \bibinfo {author} {\bibfnamefont {B.}~\bibnamefont {Zeng}}, \ and\
  \bibinfo {author} {\bibfnamefont {I.~L.}\ \bibnamefont {Chuang}},\ }\href
  {\doibase 10.1103/PhysRevA.78.062315} {\bibfield  {journal} {\bibinfo
  {journal} {Phys. Rev. A}\ }\textbf {\bibinfo {volume} {78}},\ \bibinfo
  {pages} {062315} (\bibinfo {year} {2008})}\BibitemShut {NoStop}%
\bibitem [{\citenamefont {Gheorghiu}(2011)}]{2011arXiv1101.1519G}%
  \BibitemOpen
  \bibfield  {author} {\bibinfo {author} {\bibfnamefont {V.}~\bibnamefont
  {Gheorghiu}},\ }\href@noop {} {\bibfield  {journal} {\bibinfo  {journal}
  {arXiv:1101.1519}\ } (\bibinfo {year} {2011})},\ \Eprint
  {http://arxiv.org/abs/1101.1519} {arXiv:1101.1519 [quant-ph]} \BibitemShut
  {NoStop}%
\bibitem [{\citenamefont {Gottesman}(1999)}]{tagkey19991749}%
  \BibitemOpen
  \bibfield  {author} {\bibinfo {author} {\bibfnamefont {D.}~\bibnamefont
  {Gottesman}},\ }\href {\doibase 10.1016/S0960-0779(98)00218-5} {\bibfield
  {journal} {\bibinfo  {journal} {Chaos, Solitons \& Fractals}\ }\textbf
  {\bibinfo {volume} {10}},\ \bibinfo {pages} {1749 } (\bibinfo {year}
  {1999})}\BibitemShut {NoStop}%
\bibitem [{\citenamefont {Hostens}\ \emph {et~al.}(2005)\citenamefont
  {Hostens}, \citenamefont {Dehaene},\ and\ \citenamefont
  {De~Moor}}]{PhysRevA.71.042315}%
  \BibitemOpen
  \bibfield  {author} {\bibinfo {author} {\bibfnamefont {E.}~\bibnamefont
  {Hostens}}, \bibinfo {author} {\bibfnamefont {J.}~\bibnamefont {Dehaene}}, \
  and\ \bibinfo {author} {\bibfnamefont {B.}~\bibnamefont {De~Moor}},\ }\href
  {\doibase 10.1103/PhysRevA.71.042315} {\bibfield  {journal} {\bibinfo
  {journal} {Phys. Rev. A}\ }\textbf {\bibinfo {volume} {71}},\ \bibinfo
  {pages} {042315} (\bibinfo {year} {2005})}\BibitemShut {NoStop}%
\bibitem [{\citenamefont {Jafarizadeh}\ \emph {et~al.}(2008)\citenamefont
  {Jafarizadeh}, \citenamefont {Najarbashi}, \citenamefont {Akbari},\ and\
  \citenamefont {Habibian}}]{Jafarizadeh2008EPJD}%
  \BibitemOpen
  \bibfield  {author} {\bibinfo {author} {\bibfnamefont {M.~A.}\ \bibnamefont
  {Jafarizadeh}}, \bibinfo {author} {\bibfnamefont {G.}~\bibnamefont
  {Najarbashi}}, \bibinfo {author} {\bibfnamefont {Y.}~\bibnamefont {Akbari}},
  \ and\ \bibinfo {author} {\bibfnamefont {H.}~\bibnamefont {Habibian}},\
  }\href {\doibase 10.1140/epjd/e2008-00028-0} {\bibfield  {journal} {\bibinfo
  {journal} {The European Physical Journal D}\ }\textbf {\bibinfo {volume}
  {47}},\ \bibinfo {pages} {233} (\bibinfo {year} {2008})}\BibitemShut
  {NoStop}%
\bibitem [{\citenamefont {Looi}\ and\ \citenamefont
  {Griffiths}(2011)}]{PhysRevA.84.052306}%
  \BibitemOpen
  \bibfield  {author} {\bibinfo {author} {\bibfnamefont {S.~Y.}\ \bibnamefont
  {Looi}}\ and\ \bibinfo {author} {\bibfnamefont {R.~B.}\ \bibnamefont
  {Griffiths}},\ }\href {\doibase 10.1103/PhysRevA.84.052306} {\bibfield
  {journal} {\bibinfo  {journal} {Phys. Rev. A}\ }\textbf {\bibinfo {volume}
  {84}},\ \bibinfo {pages} {052306} (\bibinfo {year} {2011})}\BibitemShut
  {NoStop}%
\end{thebibliography}%

\end{document}